\documentclass[onecolumn,letterpaper,aps,prc,longbibliography,superscriptaddress,showpacs,nofootinbib,floatfix,notitlepage]{revtex4-1}
\usepackage{lipsum}
\usepackage{graphicx,bm}
\usepackage{subcaption}
\usepackage{dblfloatfix}

\usepackage{enumerate}
\usepackage{enumitem}
\usepackage{floatrow}
\usepackage{amssymb}
\usepackage{amsmath}
\usepackage{commath}
\usepackage{verbatim}
\usepackage{xspace}
\usepackage{color}
\usepackage[%
  colorlinks=true,
  urlcolor=blue,
  linkcolor=blue,
  citecolor=blue
]{hyperref}

\usepackage{etoolbox}
\makeatletter
\patchcmd\linenumberpar{\@LN@parpgbrk}{\penalty\@LN@parpgpen\relax}{}{}
\makeatother

\newcommand{\nocontentsline}[3]{}
\newcommand{\tocless}[2]{\bgroup\let\addcontentsline=\nocontentsline#1{#2}\egroup}
\newcommand{\pp}{\ensuremath{\mathrm {p\kern-0.05em p}}}
\newcommand{\PbPb}{\ensuremath{\mbox{Pb--Pb}}}
\newcommand{\GeVc}{\ensuremath{\mathrm{GeV}\kern-0.05em/\kern-0.02em c}}
\newcommand{\sqrts}[1] {\ensuremath{\sqrt{s}=#1~\mathrm{TeV}}}
\newcommand{\pT}{\ensuremath{p_{\mathrm{T}}}}
\newcommand{\pt}{\ensuremath{p_{\mathrm{T}}}}

\newcommand{\tg}{\ensuremath{\theta_{\mathrm{g}}}}
\newcommand{\zg}{\ensuremath{z_{\mathrm{g}}}}
\newcommand{\Rg}{\ensuremath{R_{\mathrm{g}}}}
\newcommand{\zcut}{\ensuremath{z_{\mathrm{cut}}}}

\newcommand{\ptsoft}{\ensuremath{p_{\mathrm{T}}^{\mathrm{soft}}}}
\newcommand{\fmatched}{\ensuremath{f_{\mathrm{matched}}}}

\newcommand{\TeVc}{\ensuremath{\mathrm{TeV}\kern-0.05em/\kern-0.02em c}}

\newcommand{\com}[1]       {}

\newcommand{\RAA}          {\ensuremath{R_{\mathrm{AA}}}}

\newcommand{\akT}          {\ensuremath{{\rm anti-}k_{\rm T}}}

\usepackage{geometry}
\begin{document}
%
%
%
\title{Identifying groomed jet splittings in heavy-ion collisions}

\newcommand{\ucb}{Physics Department, University of California, Berkeley, Berkeley, CA, USA}
\newcommand{\lbl}{Nuclear Science Division, Lawrence Berkeley National Laboratory, Berkeley, CA, USA}

\author{James Mulligan} 
\affiliation{\ucb}
\affiliation{\lbl}
\author{Mateusz P\l osko\'n} 
\affiliation{\lbl}

\date{October 28 2020}

\begin{abstract}

Measurements of jet substructure in heavy-ion collisions may
provide key insight into the nature of jet quenching in the quark-gluon plasma.
Jet grooming techniques from high-energy physics have been applied to heavy-ion
collisions in order to isolate theoretically controlled jet observables and
explore possible modification to the hard substructure of jets.
However, the grooming algorithms used have not been tailored to the unique
considerations of heavy-ion collisions, 
in particular to the experimental challenge of reconstructing jets in the presence
of a large underlying event.
We report a set of simple studies illustrating 
the impact of the underlying event on identifying groomed jet 
splittings in heavy-ion collisions, and on associated 
groomed jet observables.
We illustrate the importance of the selection of the grooming algorithm,
as certain groomers are more robust against these effects, while others, including
those commonly used in heavy-ion collisions, are susceptible to large
background effects -- which, when uncontrolled, can mimic a jet quenching signal.
These experimental considerations, along with appropriate theoretical motivation, 
provide input to the choice of grooming algorithms employed in 
heavy-ion collisions.

\end{abstract}

\maketitle




\section{Introduction}

Jet grooming techniques were developed in the high-energy
physics community to mitigate pileup contamination and improve the
theoretical calculability of jet observables in \pp{} collisions.
The Soft Drop algorithm, for example, reduces non-perturbative effects 
by selectively removing soft large-angle radiation,
which allows for well-controlled comparisons of measurements with perturbative QCD (pQCD) calculations
\cite{Larkoski:2014wba, Dasgupta:2013ihk, Larkoski:2015lea}.
Grooming techniques have recently been applied to heavy-ion collisions
in order to establish whether jet quenching in the quark-gluon plasma 
modifies the hard substructure of jets, such as the splitting function,
and to elucidate whether jets lose energy coherently, as a single color charge, 
or incoherently, as multiple independent substructures
\cite{PhysRevLett.119.112301, Mehtar-Tani2017, CHANG2018423, JEWEL2017, Elayavalli2017, Caucal2019, Casalderrey-Solana2020, Ringer:2019rfk, Andrews_2020}.
Moreover, Monte Carlo (MC) event generators suggest that 
jet splittings identified by grooming algorithms
are correlated to parton shower splittings,
raising the possibility that identifying groomed jet splittings
in heavy-ion collisions may provide a handle on the space-time
evolution of jet propagation through the hot QCD medium.

Measurements of the Soft Drop groomed momentum fraction \zg{}
have been made in \pp{} and heavy-ion collisions at the Large Hadron Collider (LHC) 
and the Relativistic Heavy Ion Collider (RHIC)
\cite{PhysRevD.101.052007, collaboration2020measurement, PhysRevLett.120.142302, 2020135227, kauder2017measurement}.
These measurements have opened a new avenue in heavy-ion jet physics.
Measurements by the CMS and ALICE Collaborations show a modification
of the \zg{} distribution in \PbPb{} collisions relative to \pp{} collisions
-- however, the results have not been corrected for background effects. 
Local background fluctuations in a heavy-ion
environment can result in an incorrect splitting (unrelated to the jet) 
being identified by the grooming algorithm. 
This problem is analogous to the well-known
experimental problem of ``combinatorial'' jets in heavy-ion collisions,
which is typically treated by either (1) Reporting jet measurements
in the background-free region of phase space, namely at sufficiently
large \pT{} and/or small $R$, or (2) Subtracting the combinatorial
jet distribution on an ensemble basis. 
In the case of groomed jet observables, the scale at which background 
effects occur is set by the subleading prong of the groomed jet, 
rather than the jet \pT{} and $R$.
The presence of background contamination in groomed jet observables 
has been recognized to some extent since the first measurements in heavy-ion 
collisions; however, the magnitude of the effect has not been quantified, nor
has its qualitative impact been understood. Since the reported
distributions contain a significant number of ``mistagged'' splittings,
it remains unclear how to interpret the observed modifications.

Since the characteristic scale of these effects is set by the subleading 
prong of the groomed jet, the impact of local background fluctuations
on groomed jet observables depends on the grooming algorithm employed. 
In this article, we present a simple set of studies on the
performance of various grooming algorithms with respect to background
contamination effects in heavy-ion collisions, in order to confront 
the experimental question: How are grooming algorithms affected by 
the presence of a heavy-ion background?
We identify groomers that are relatively robust to background effects, as
well as those that are susceptible to contamination.
Finally, we discuss implications on the interpretation 
of previous measurements.

\section{Analysis Setup}

We reconstruct jets from charged particles 
in central rapidity generated by PYTHIA \cite{Sjostrand:2014zea} 
for proton-proton collisions at \sqrts{5}
using the \akT\ algorithm from the FASTJET \cite{antikt} 
package with resolution parameter $R=0.4$. 
Before the jet finding 
we select particles with $p_{T} > 0.15 \;\GeVc$. 
This setup corresponds to typical experimental configurations at the LHC.
To approximate the heavy-ion background, we use a thermal model
consisting of $N$ particles drawn from a Gaussian with 
$\left< \frac{dN}{d\eta}\right> \approx 1800 $ and \pT{} sampled from a Gamma distribution: 
$f_\Gamma \left( \pT;\alpha,\beta \right) \propto \pT^{\alpha-1} e^{-\pT/\beta}$
with $\alpha=2$. We select $\beta=0.5$ in order to roughly fit 
the width of the $R=0.4$ $\delta \pT{}$ distribution in 0-10\% \PbPb{} data 
of $\sigma \approx 11 \;\GeVc$ \cite{Abelev2012}.
We perform event-wide constituent subtraction on
the combined event consisting of the charged particles from the 
PYTHIA event together with the thermal background particles,
using $R_{\mathrm{max}}=0.25$ \cite{Berta2019}.
We then cluster the subtracted particles into jets, 
and match these ``combined'' jets to
those jets found by clustering only the PYTHIA particles.

\subsection{Groomers}

To study the performance of different grooming criteria,
we use the Soft Drop algorithm \cite{Larkoski:2014wba} and 
the Dynamical Grooming algorithm
\cite{PhysRevD.101.034004,mehtartani2020tagging} but also 
new rather simple groomers which we call max-$z$, max-$p_{T}^{\mathrm{soft}}$, 
max-$\kappa$, max-$k_{T}$, and min-$t_{f}$. These are all defined by
reclustering the jet with the Cambridge/Aachen (CA) algorithm,
where every step of the clustering history is defined by a radiator and two 
prongs that it decays to.
We denote the two prongs 
$a$ and $b$ such that $p_{T}^{\mathrm{radiator}}=p_{T}^{a} + p_{T}^{b}$, where
$p_{T}^{b} < p_{T}^{a}$, and $R_{g}=\sqrt{(y_{a}-y_{b})^{2} + (\varphi_{a} - \varphi_{b})^{2}}$ is the angular separation between the two 
(used interchangeably with $\tg \equiv R_{g}/R$) with $\varphi$ 
being the azimuthal angle and $y$ the rapidity of the prongs. 
Therefore, $k_{T} \equiv p_{T}^{b}R_{g}$, 
$z \equiv p_{T}^{b}/p_{T}^{\mathrm{radiator}}$, 
and $\kappa \equiv zR_{g}$.
We briefly describe the algorithms that we use below:

\begin{enumerate}[label=(\roman*)]
    \itemsep0em
    \item Soft Drop with $\beta=0$ with three values of the symmetry parameter $\zcut=0.1, 0.2, 0.3$.
    \item Dynamical Grooming with three values of the grooming parameter $a=0.1, 1.0, 2.0$.
    \item max-$z$: For every jet that contains more than one particle, identify
    the splitting where $z$ is the largest from all the splittings in the
    primary Lund plane.
    \item max-$p_{T}^{\mathrm{soft}}$: For every jet that contains more than one
    particle, identify the splitting where the soft prong has the largest \pt{} 
    from all of the softer prongs within any pair in the primary Lund plane. 
    \item max-$\kappa$: For every jet that contains more than one particle, 
    identify the splitting where $\kappa$ is the largest from all splittings in the primary Lund plane.
    \item max-$k_{T}$: For every jet that contains more than one particle,
    identify the splitting where $k_{T}$ is the largest from all splittings 
    in the primary Lund plane.
    \item min-$t_{f}$: For every jet that contains more than one particle, 
    identify the splitting where $zR_{g}^{2}$ is the largest from all
    the splittings in the primary Lund plane (in relation to the estimate 
    of the formation time for the pair $t_{f} \sim \frac{1}{zR_{g}^{2}}$).
\end{enumerate}

\begin{figure}

\begin{subfigure}{\linewidth}
\includegraphics[width=0.95\textwidth]{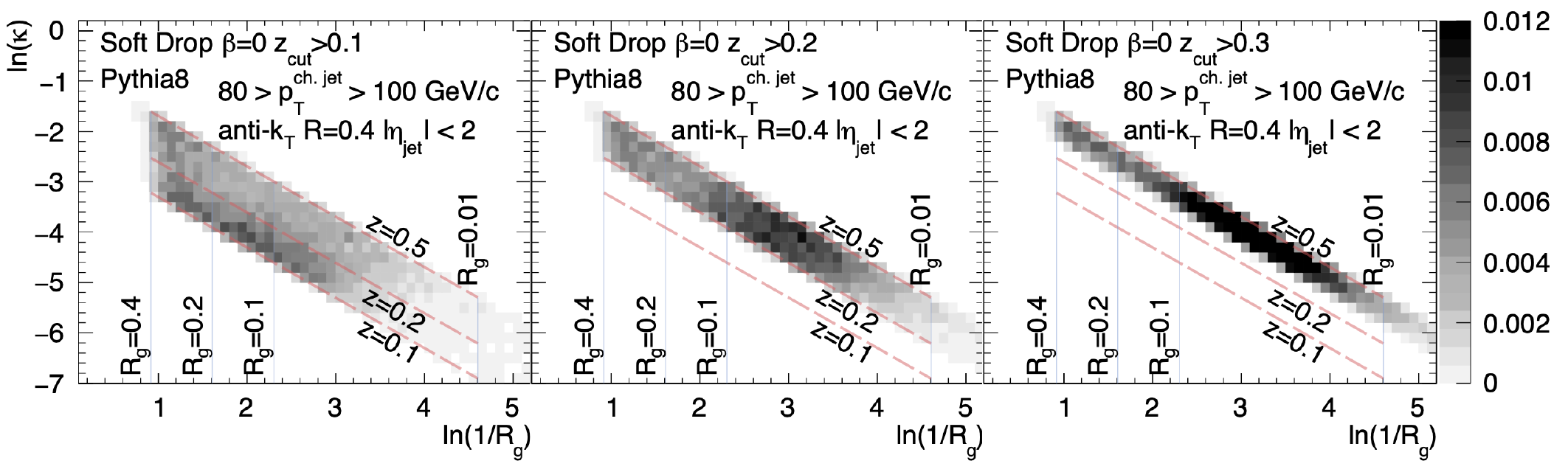}
\caption{Primary Lund plane obtained with Soft Drop grooming with $\beta=0$ for different symmetry cut \zcut{} parameters. Left: $\zcut = 0.1$. Middle: $\zcut = 0.2$. Right: $\zcut = 0.3$.}
\end{subfigure}

\begin{subfigure}{\linewidth}
\includegraphics[width=0.95\textwidth]{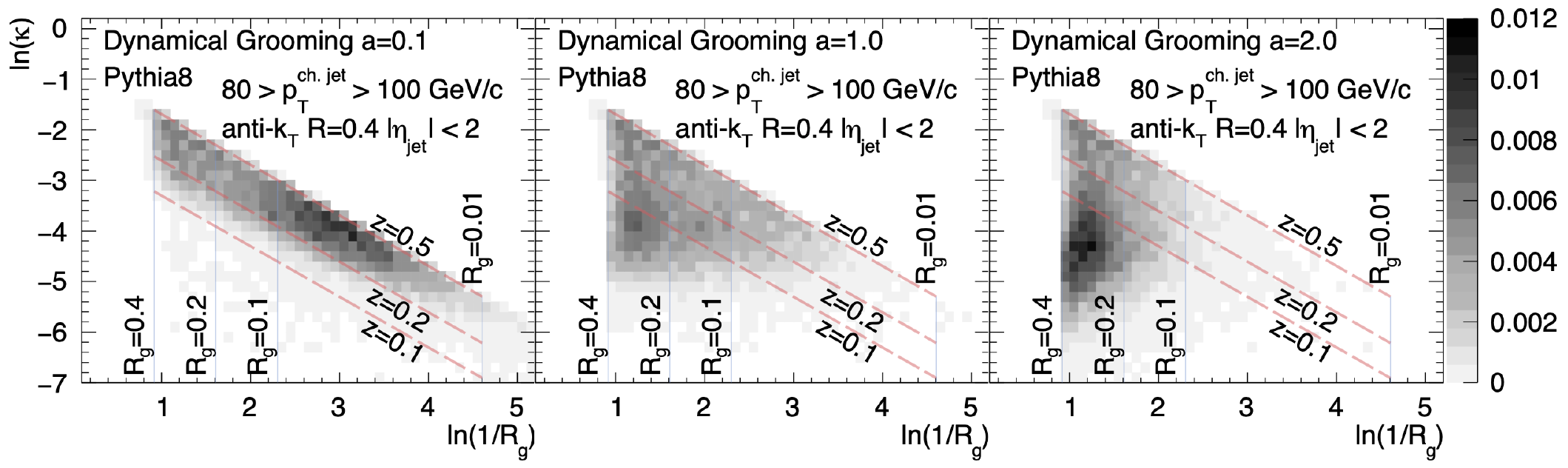}
\caption{Primary Lund plane obtained with Dynamical Grooming for different 
values of $a$. Left: $a=0.1$. Middle: $a=1.0$. Right: $a=2.0$. }
\end{subfigure}

\begin{subfigure}{\linewidth}
\includegraphics[width=0.6666\textwidth]{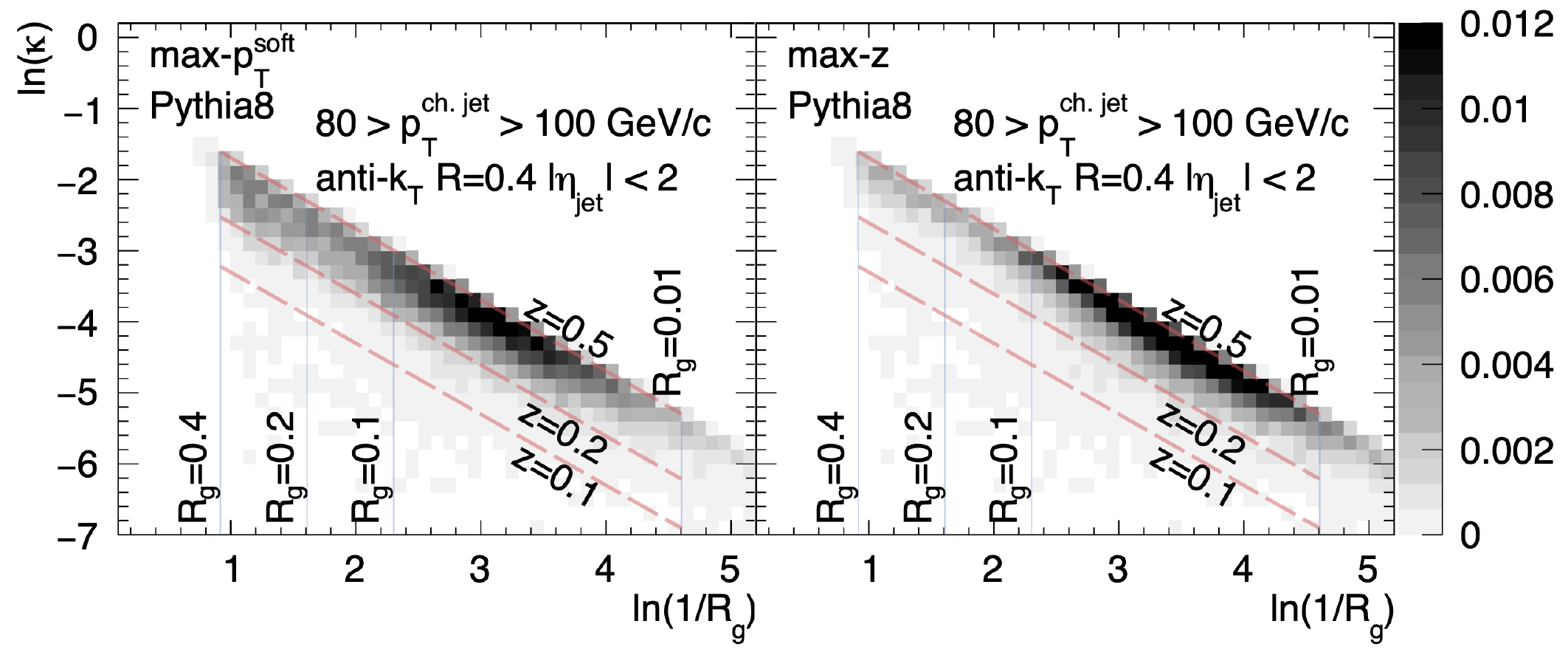}
\caption{Primary Lund plane obtained with new groomers with the split selection depending on momentum of the prongs. Left: max-\ptsoft. Right: max-$z$. }
\end{subfigure}

\begin{subfigure}{\linewidth}
\includegraphics[width=0.95\textwidth]{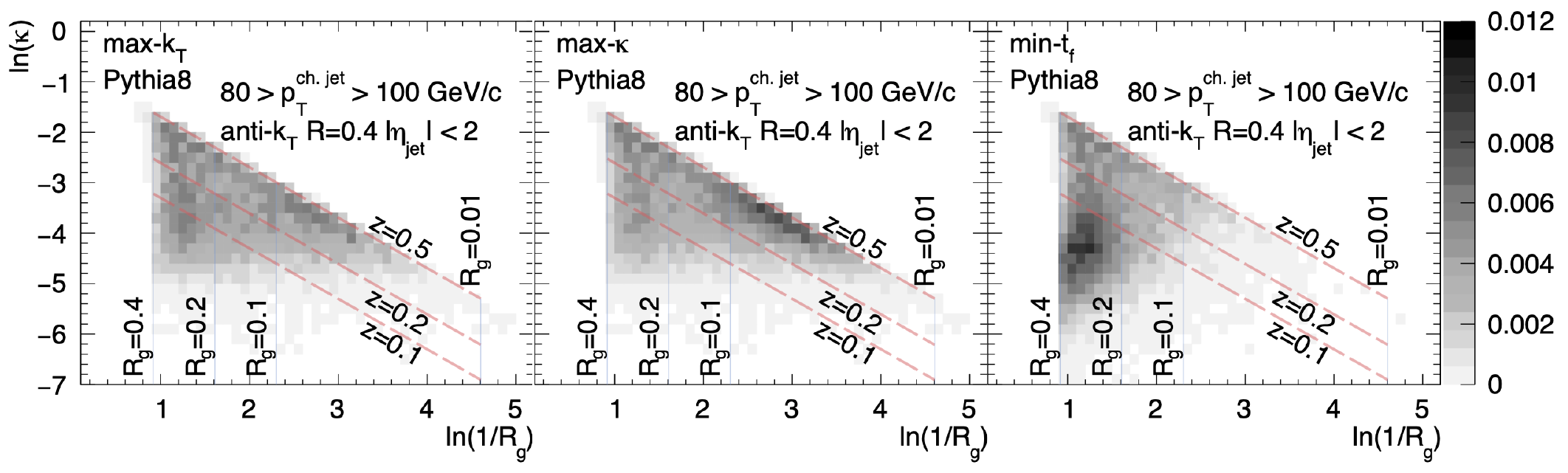}
\caption{Primary Lund plane obtained with new groomers with the split selection depending on momentum and the angle between the prongs. Left: max-$k_{T}$. Middle: max-$\kappa$. Right: min-$t_{f}$.}
\end{subfigure}

\caption{Primary Lund plane density diagram of groomed splittings for various groomers. Events generated using PYTHIA for proton-proton collisions at \sqrts{5}. Jets reconstructed from charged particles at hadron-level.}

\label{fig2}
\end{figure}

For an overview of the phase space that each of the grooming algorithms selects,
we plot the primary Lund plane density 
$\rho(\kappa,R_{g})= \frac{1}{N_{jet}} \frac{\mathrm{d}^{2}N}{\mathrm{d}\ln(\kappa) / \mathrm{d}\ln(1/R_{g})}$
for identified splittings in Fig. \ref{fig2} \cite{Dreyer2018}.
We note that several of these groomers are expected to select similar phase space:
max-$z$, max-$p_{T}^{\mathrm{soft}}$, and Dynamical Grooming $a=0.1$ select 
approximately on the longitudinal momentum of the splitting; 
max-$\kappa$, max-$k_{T}$, and Dynamical Grooming $a=1.0$ select 
approximately on the transverse momentum of the splitting; 
min-$t_{f}$ and Dynamical Grooming $a=2.0$ select 
approximately on the mass of the splitting.

\subsection{Prong matching} \label{prong-matching}

In order to study the impact of the heavy-ion background on the reconstruction of
groomed splittings, we examine where $>50\%$ of the PYTHIA subleading 
prong (by \pt{}) is reconstructed in the combined event. 
We consider only the case where both the PYTHIA jet and the
combined jet pass the grooming condition. We categorize six possibilities -- the
PYTHIA subleading prong haas one of the following characteristics:

\begin{figure}[!t]
\centering{}
\includegraphics[scale=0.55]{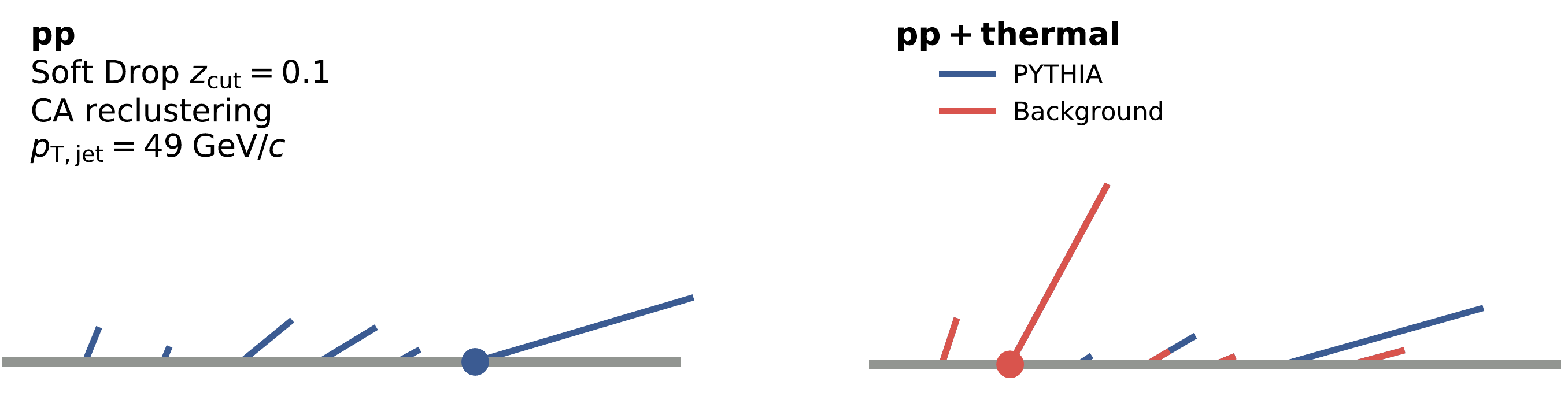}
\caption{Example of a PYTHIA jet (left) and the same jet embedded into
thermal background (right). 
In the case of thermal background,
a background fluctuation at large angle passing
the grooming condition results in the subleading prong being 
absorbed in the leading prong.}
\label{fig1}
\end{figure}

\begin{enumerate}[label=(\arabic*)]
    \itemsep0em
    \item It is correctly reconstructed in the subleading prong of the combined jet.
    \item It is reconstructed in the leading prong of the
    combined jet, and the PYTHIA leading prong is reconstructed in the subleading
    prong of the combined jet. That is, both prongs are correctly identified, but 
    they ``swap'' which is leading and which is subleading.
    In this case, \zg{} and \tg{} are invariant --
    although iterative observables are not.
    \item It is reconstructed in the leading prong of the
    combined event, and the PYTHIA leading prong is not reconstructed in the subleading
    prong of the combined event. This is the most common way that an incorrect splitting
    is reconstructed, typically by a background fluctuation at large angle passing
    the grooming condition. Due to angular clustering, this by definition results in
    the subleading prong being absorbed in the leading prong, as shown in
    Fig. \ref{fig1}.
    \item It is reconstructed in the groomed-away constituents of the combined jet.
    \item It is reconstructed nowhere in the combined jet, but 
    rather its constituents are elsewhere in the combined event.
    \item It is not reconstructed in any of the above
    categories; for example, it may have 1/3 of its \pt{} split between three
    categories.
\end{enumerate}

\section{Performance of groomers}

\begin{figure}[!b]
\centering{}
\includegraphics[scale=0.55]{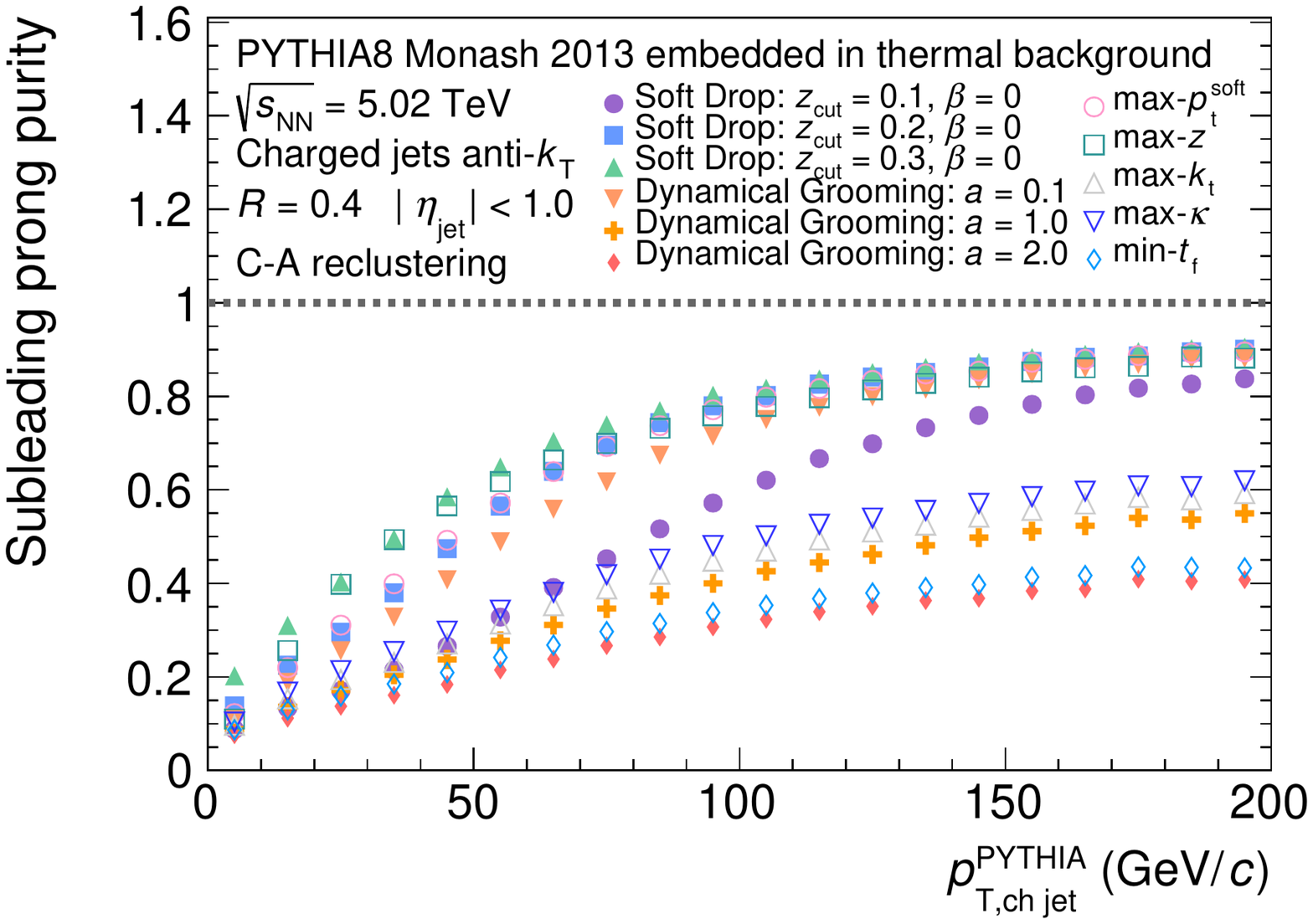}
\caption{Subleading prong purity as a function of \pt{} for a variety
of groomers. Note that the purity defined here includes only those cases where
the PYTHIA subleading prong is correctly tagged as the subleading prong in the
combined jet.}
\label{fig3}
\end{figure}

For each groomer, we plot the fraction of subleading prongs in the combined
events that are correctly tagged in Figure \ref{fig3}, as a function 
of jet \pt{}.
Immediately, it is apparent that to increase the subleading prong purity one 
should (i) Choose a suitable groomer, and/or (ii) Measure high-\pt{} jets.
Groomers with an angular selection perform the worst, which is unsurprising
given that combinatorial background preferentially occupies large-angle phase space,
as compared to jets.
Groomers which select on longitudinal momentum (Dynamical grooming $a=0.1$, 
max-\ptsoft{}, max-$z$) perform well, with Dynamical grooming performing
slightly worse, presumably due to its small angular component in the grooming condition.
Soft Drop performs similarly to these for $\zcut = 0.2,0.3$, where above 
$\pt=70 \;\GeVc$ there
appears to be an approximate saturation, in which case further increasing \zcut{} does 
not increase the purity. Soft Drop with $\zcut = 0.1$, which is the most common
configuration used in heavy-ion collisions, performs notably worse.
This suggests that mistagged splittings arise from a characteristic longitudinal 
momentum scale above which background is suppressed, due to uncorrelated background 
fluctuations on the geometric scale of a prong.
These results were repeated using Angantyr \cite{Bierlich2018} 
to model the underlying heavy-ion event,
and similar results were obtained, with identical ordering of the groomers and 
purities within approximately 10\% compared to the thermal background.

\begin{figure}[!h]
\centering{}
\includegraphics[scale=0.55]{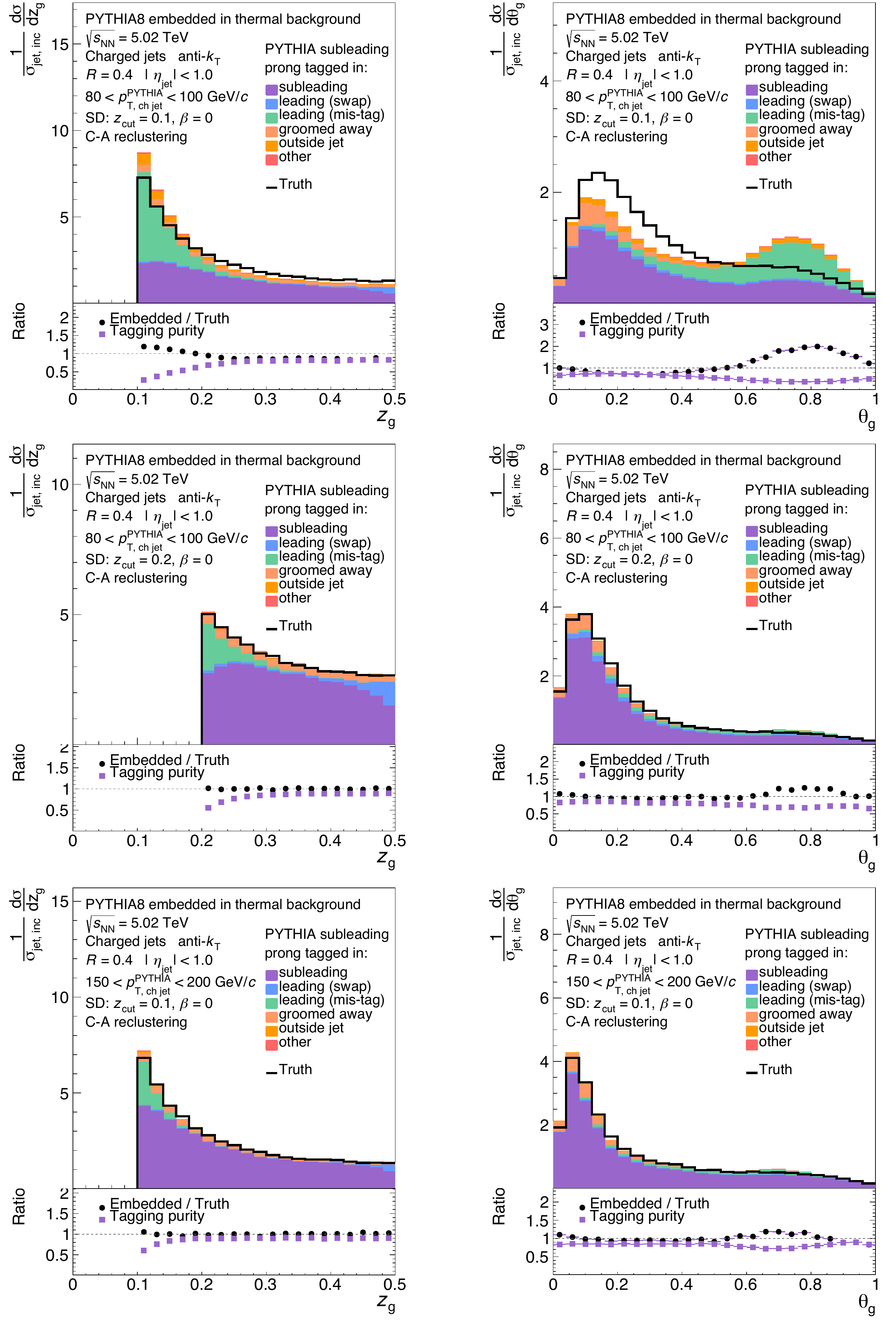}
\caption{
Distributions of \zg{} (left) and \tg{} (right) when PYTHIA is embedded in the 
heavy-ion background, as well as from PYTHIA (``Truth'').
The bottom panels show the purity and the ratio of the embedded distribution to the
PYTHIA distribution.
Top: Low-\pt{}, $\zcut=0.1$.
Middle: Low-\pt{}, $\zcut=0.2$.
Bottom:  High-\pt{}, $\zcut=0.1$.}
\label{fig:money-plots}
\end{figure}

To determine the dependence of the mistagged fraction on the
splitting observables, we decompose the distributions of \zg{}, \tg{} 
according to where the PYTHIA subleading prong is reconstructed in 
the combined event, as described in Section \ref{prong-matching}.
Figure \ref{fig:money-plots} shows the \zg{} (left) and \tg{} (right) distributions 
when PYTHIA is embedded in the heavy-ion background. 
For smaller \zcut{} and lower \pt{} (top row),
there is a large fraction of mistagged splittings, predominantly from the case
where the subleading prong is mistagged in the leading prong (Fig. \ref{fig1}).
The mistagged prongs are most prominent at small $z$ 
(where the true \zg{} distribution is naturally peaked) 
and large $\theta$ (in the tail of the true \tg{} distribution);
however, they are not limited to these regions of phase space. 
We note that the correctly tagged distributions
exhibit significant deviations from the true
distributions, suggesting that there are strong correlations between the 
structure of the jet and its susceptibility to mistagging.
By raising \zcut{} (middle row) or increasing \pt{} (bottom row), 
the mistagging rates 
are significantly reduced -- suggesting that at low-\pt{}, the Soft Drop
groomer with $\zcut=0.1$ is undesirable in heavy-ion collisions, and even with
larger \zcut{} or higher \pt{} one should proceed with caution.
The bottom panels of Fig. \ref{fig:money-plots} show the fraction of subleading prongs
in the embedded events that are correctly tagged, 
which is denoted as tagging purity [where we now include cases (1) and (2)
from Section \ref{prong-matching} as correct identification].
We additionally plot the ratio of the embedded distribution to the true distribution, 
which shows significant deviations, typically larger for \tg{} than \zg{}.

\begin{figure}[!b]
\centering{}
\includegraphics[scale=0.45]{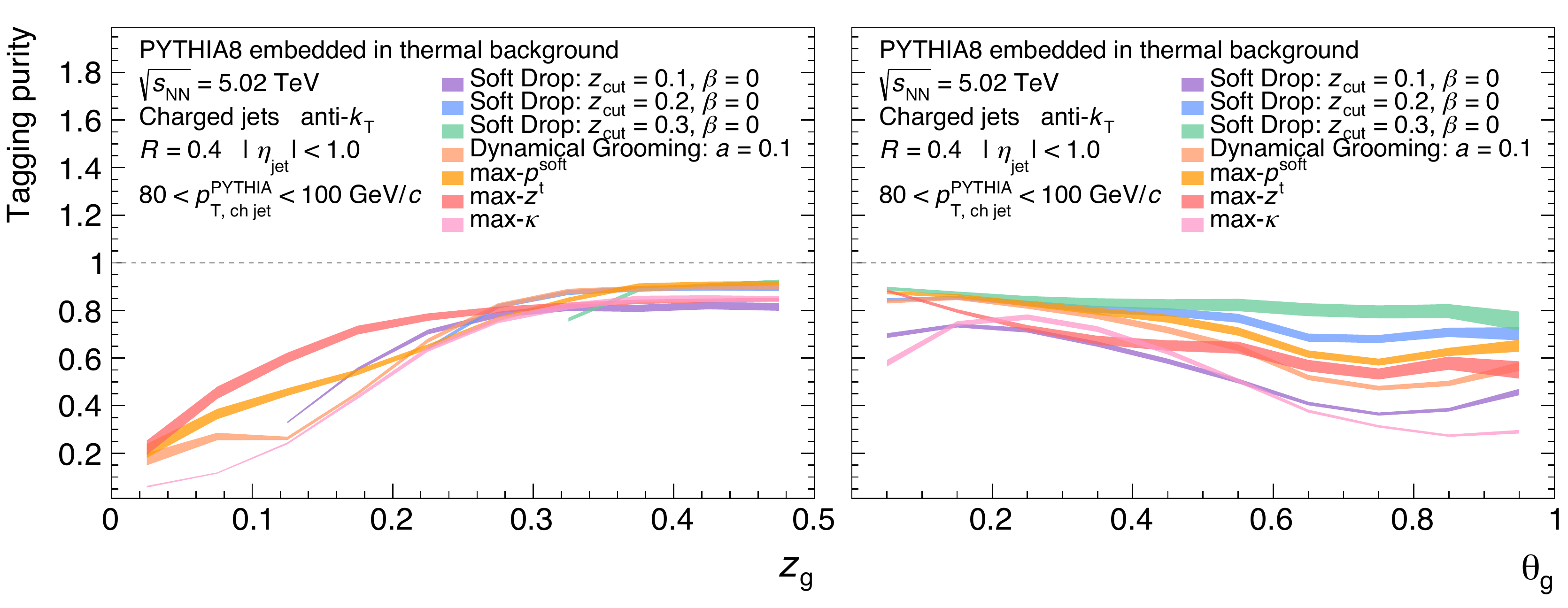}
\caption{Subleading prong purities as a function of \zg{} (left) and \tg{} (right)
for a variety of groomers.}
\label{fig:ratios-purity}
\end{figure}

To investigate the robustness of the choice of grooming algorithm
to these experimental background effects, we plot the two
ratios from the bottom panels of Fig. \ref{fig:money-plots} for a variety of groomers.
In Fig. \ref{fig:ratios-purity}, we plot the subleading prong tagging purity. 
For \zg{} (left), the purity is high at large \zg{}, but decreases substantially at small \zg{}.
For \tg{} (right), on the other hand, the purity is typically highest at low \tg{}, and decreases
at large \tg{}. Groomers which select on the longitudinal hardness of the splitting
(Soft Drop, Dynamical Grooming $a=0.1$, max-\ptsoft{}, and max-$z$) perform the best;
however, even in these cases the purity becomes low when the absolute scale of $z$
becomes small (Soft Drop $\zcut=0.1$, and all others for \zg{} small). 
Of the groomers considered here, Soft Drop is the only one with an absolute cutoff in the
grooming condition, which constrains the observable to the high-purity region.
This, in combination with the well-studied theoretical benefits of Soft Drop, 
suggests that Soft Drop with sufficiently large \zcut{} is an appealing groomer
for heavy-ion collisions.
We note however that in this \pt{} range, the purity remains significantly
less than unity, which must be treated carefully.
Nevertheless, by maximizing the purity, one can achieve improved
experimental control, both by reducing the magnitude of corrections and modeling
needed in the measurement, but also by enabling a stable unfolding procedure
due to the rejection of large off-diagonal contamination of the response matrix,
which is otherwise often unfeasible.

\begin{figure}[!t]
\centering{}
\includegraphics[scale=0.45]{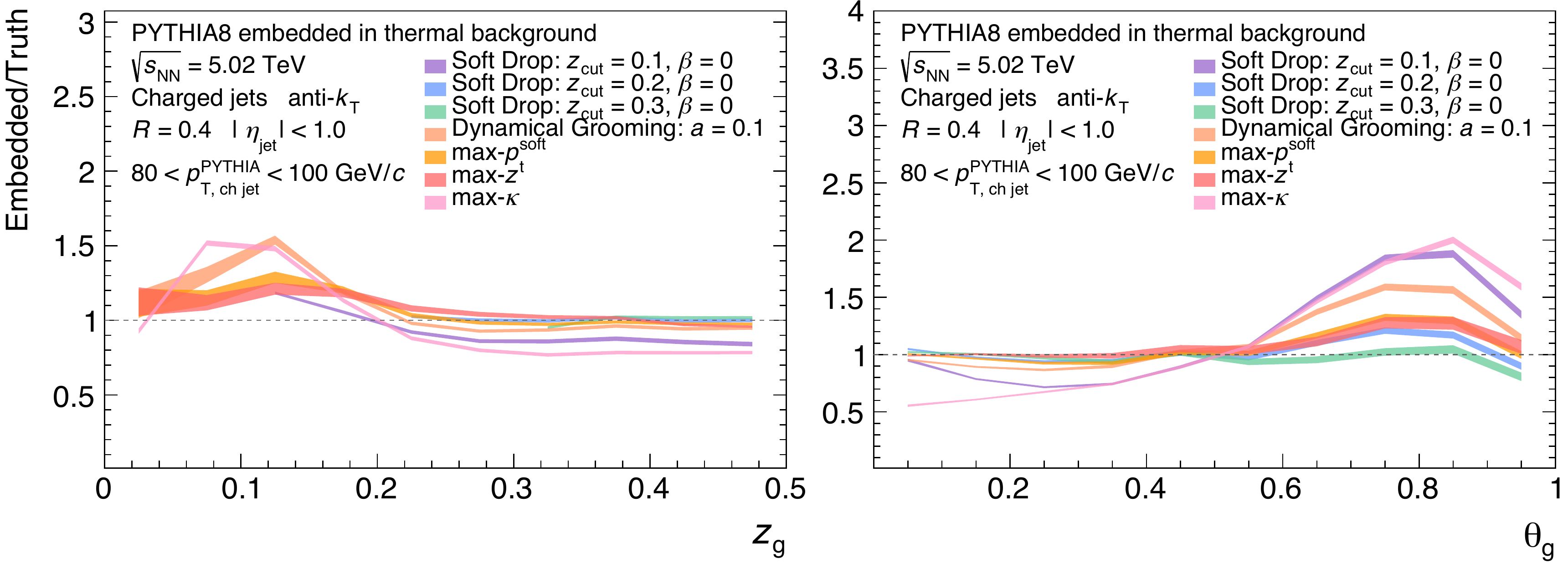}
\caption{Ratio of embedded \zg{} (left) and \tg{} (right) distributions to PYTHIA 
for a variety of groomers.}
\label{fig:ratios-groomers}
\end{figure}

Figure \ref{fig:ratios-groomers} shows ratio of the embedded \zg{}
and \tg{} distributions 
to the PYTHIA distributions for a variety of groomers.
This provides complementary information to the purity, since it describes
the impact not only of the fraction of mistagged splittings, but how different
the mistagged splittings are from the true splittings.
Similar to the purity, the Soft Drop $\zcut=0.1$ and max-$\kappa$ groomers perform
poorly, whereas the other groomers perform relatively well.
We see that this ratio is typically nearer to unity for \zg{} compared to \tg{},
since for \zg{} the mistagged splittings typically deplete and re-populate the low-$z$ region, 
whereas for \tg{} the mistagged splittings are likely to populate large angles.

\begin{figure}[!h]
\centering{}
\includegraphics[scale=0.33]{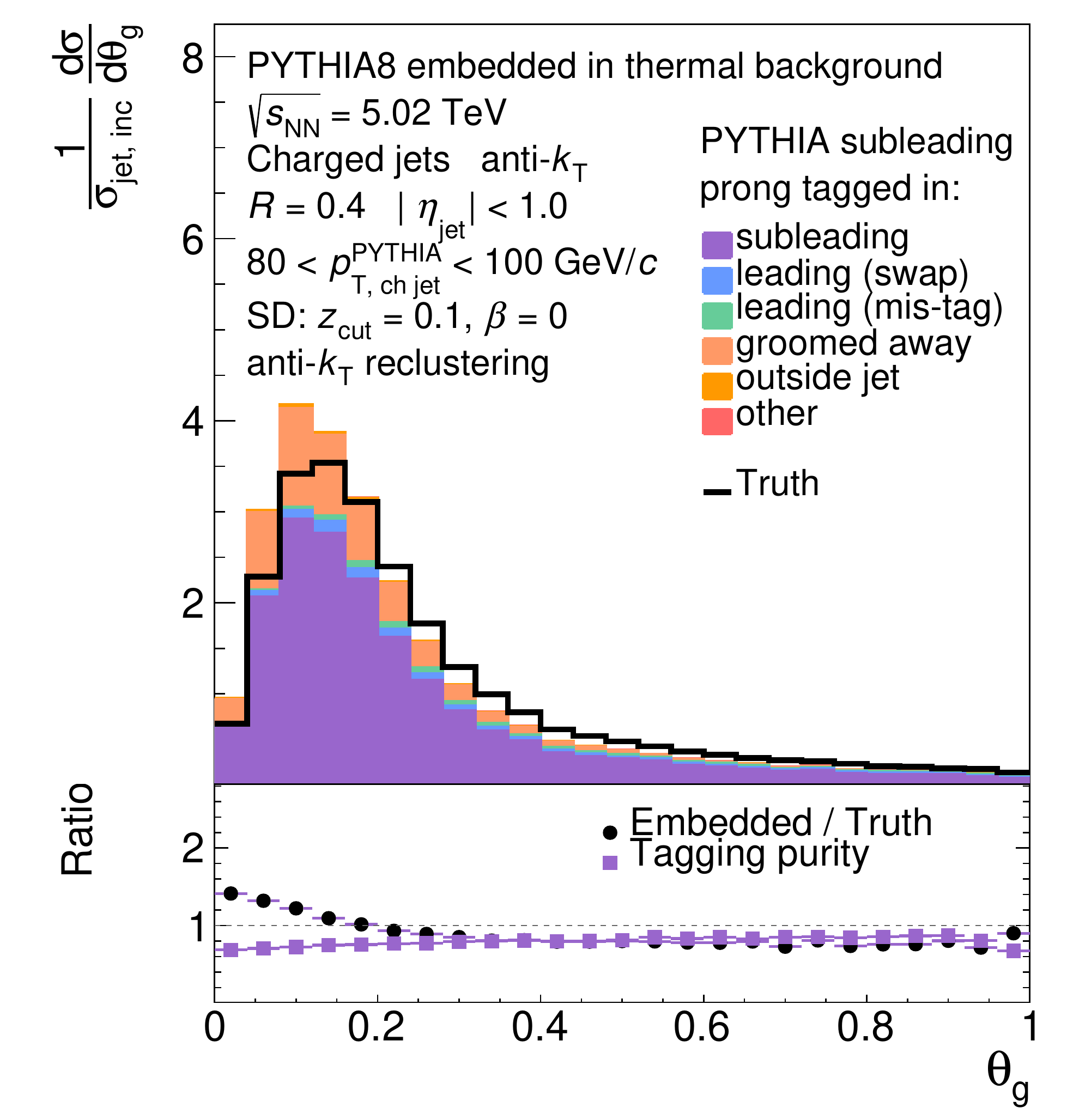}
\includegraphics[scale=0.33]{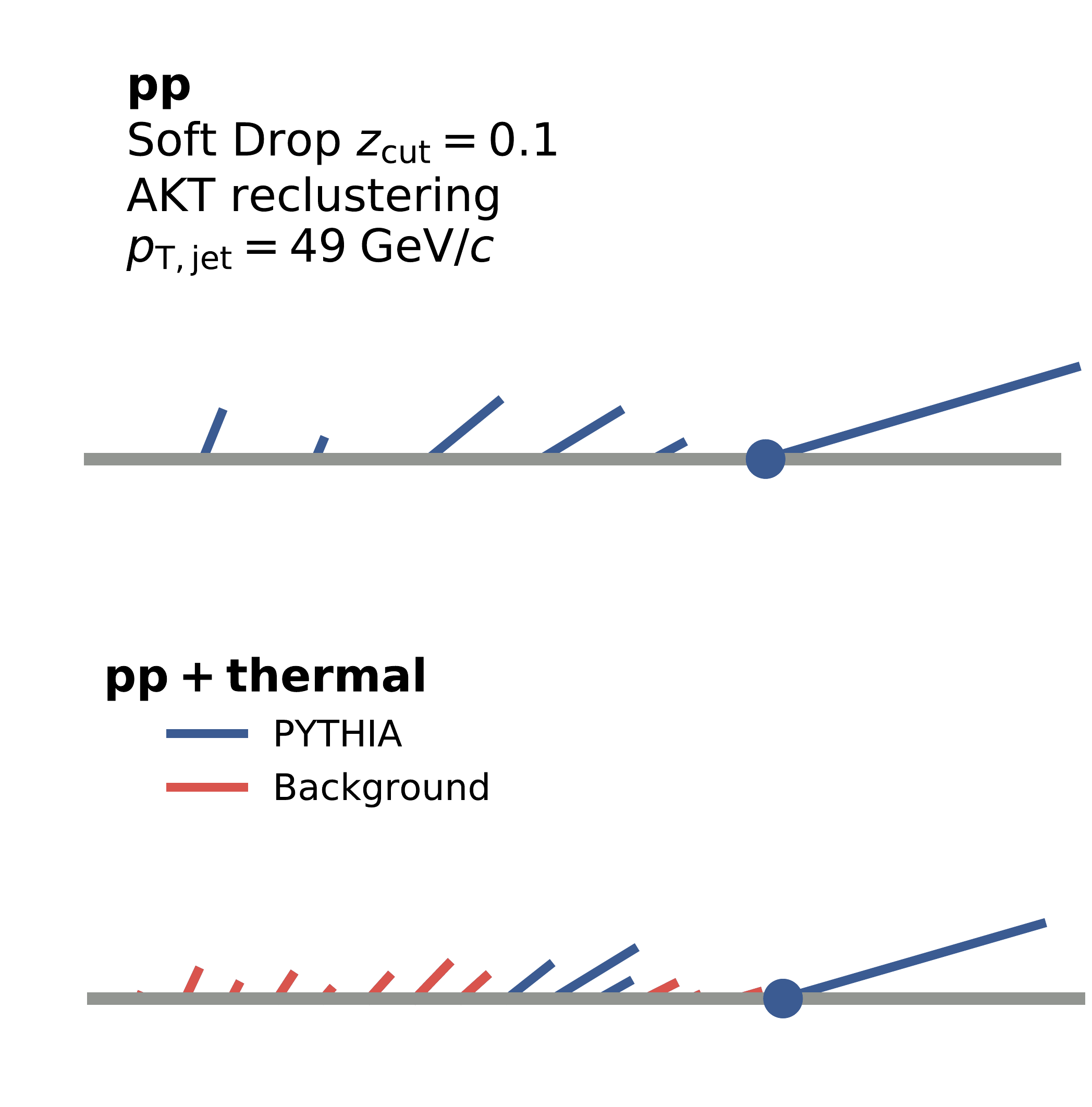}
\caption{
Left: Distributions of \tg{} when PYTHIA is embedded in the 
heavy-ion background, as well as from PYTHIA (``Truth''), 
for anti-$k_{\mathrm{T}}$ reclustering.
Right: Example of a PYTHIA jet (identical to Fig. \ref{fig1}) 
and the same jet embedded into thermal background for 
anti-$k_{\mathrm{T}}$ reclustering. 
In the case of thermal background,
anti-$k_{\mathrm{T}}$ reclustering results in large-angle background particles being
individually clustered to the leading branch, which results in them being groomed away.
}
\label{fig:akt}
\end{figure}

Finally, we note that the choice of reclustering algorithm can have a large impact
on the splitting purity. To illustrate this, in Fig. \ref{fig:akt} (left)
we plot the mistagging distribution as a function of \tg{} 
for Soft Drop $\zcut=0.1$ with anti-$k_{\mathrm{T}}$ (AKT) reclustering.
Compared to CA reclustering (Fig. \ref{fig:money-plots} top right),
the purity is improved by approximately 20\% at intermediate \pT, 
and the large-angle mistagging is absent.
This behavior can be understood since the anti-$k_{\mathrm{T}}$ reclustering 
sequence is fundamentally different than that of CA.
The anti-$k_{\mathrm{T}}$ algorithm tends to cluster branches where at least one branch
has large \pT{}, resulting in a clustering sequence
dominated by the leading prong clustering together with surrounding individual particles
-- as compared with CA, 
which allows softer particles to cluster among themselves before combining with
the leading prong.
Accordingly, anti-$k_{\mathrm{T}}$ reclustering has on average a larger number of primary splittings compared with CA reclustering.
Since background prongs typically arise from local
fluctuations of particle number at large angle, anti-$k_{\mathrm{T}}$ reclustering 
leads to enhancement of 
the purity as large-angle particles are individually clustered and then groomed away.
This is illustrated in Fig. \ref{fig:akt} (right) for the same PYTHIA jet
as in Fig. \ref{fig1}.
However, it is important to note that anti-$k_{\mathrm{T}}$ reclustering 
has certain theoretical drawbacks \cite{Frye2016}, and may therefore be undesirable.
Nevertheless, due to the observed benefits with regard to background contamination, 
it may be worth further theoretical consideration.

\section{Relevance to previous measurements}

In this section, we briefly outline the implications of our studies
on the interpretation of published measurements of \zg{} 
\cite{2020135227,PhysRevLett.120.142302}.
These measurements are reported without corrections for background effects
or detector effects, but rather \PbPb{} data is compared with an embedded reference.
In both \cite{2020135227,PhysRevLett.120.142302}, cuts on \Rg{} are employed, which are expected to induce
suppression (or enhancement) of the remaining \zg{} distribution
in \PbPb{} relative to pp.\footnote{The measurements are normalized differently: 
In the case of the CMS Collaboration, any suppression due to the \Rg{} cut is self-normalized away, 
whereas in the case of the ALICE Collaboration, any suppression due to the \Rg{} cut persists in 
the \zg{} distribution.}
There are two relevant effects that the presence of mistagged splittings 
can have on such measurements.

\begin{figure}[!t]
\centering{}
\includegraphics[scale=0.35]{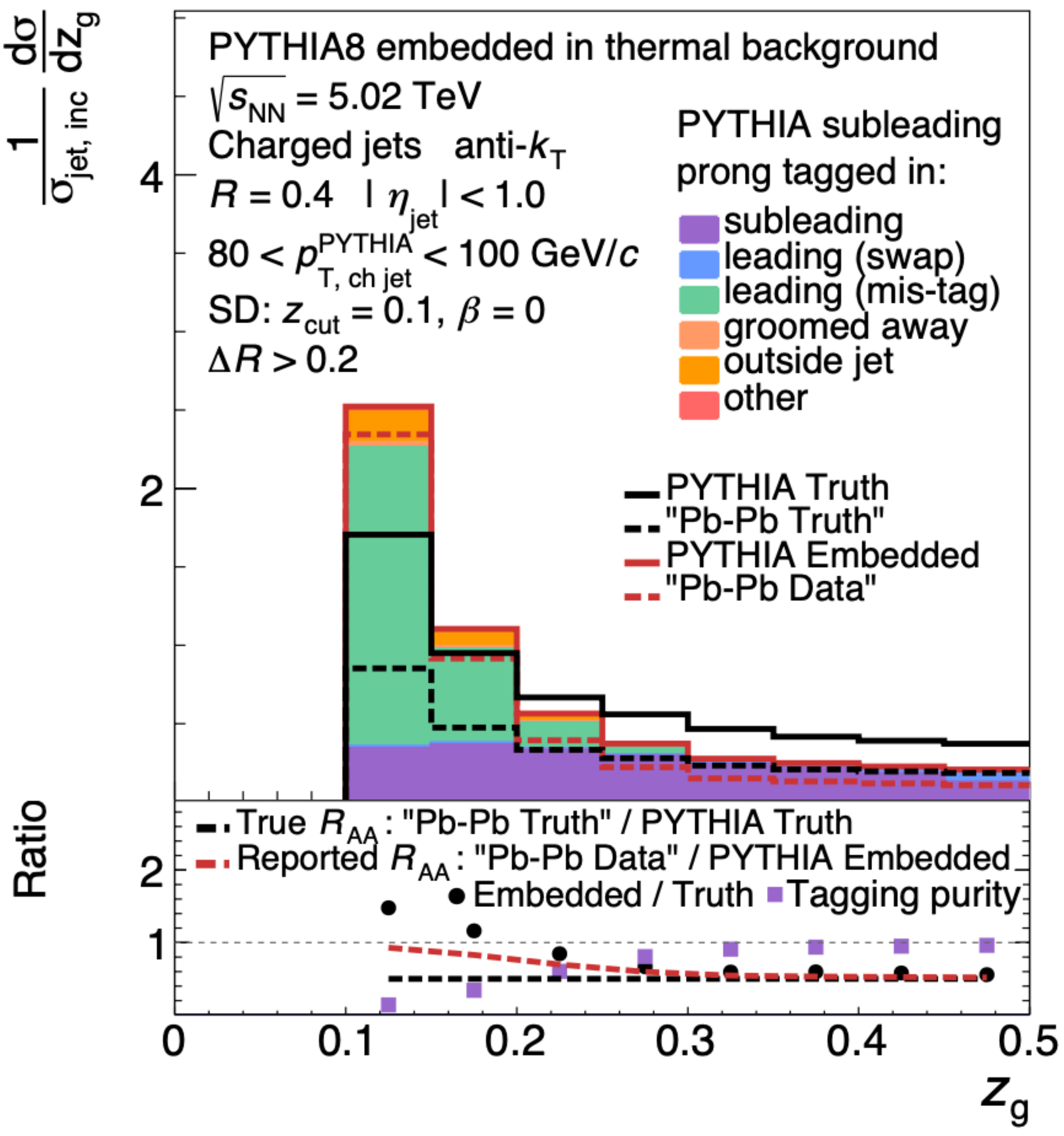}
\includegraphics[scale=0.35]{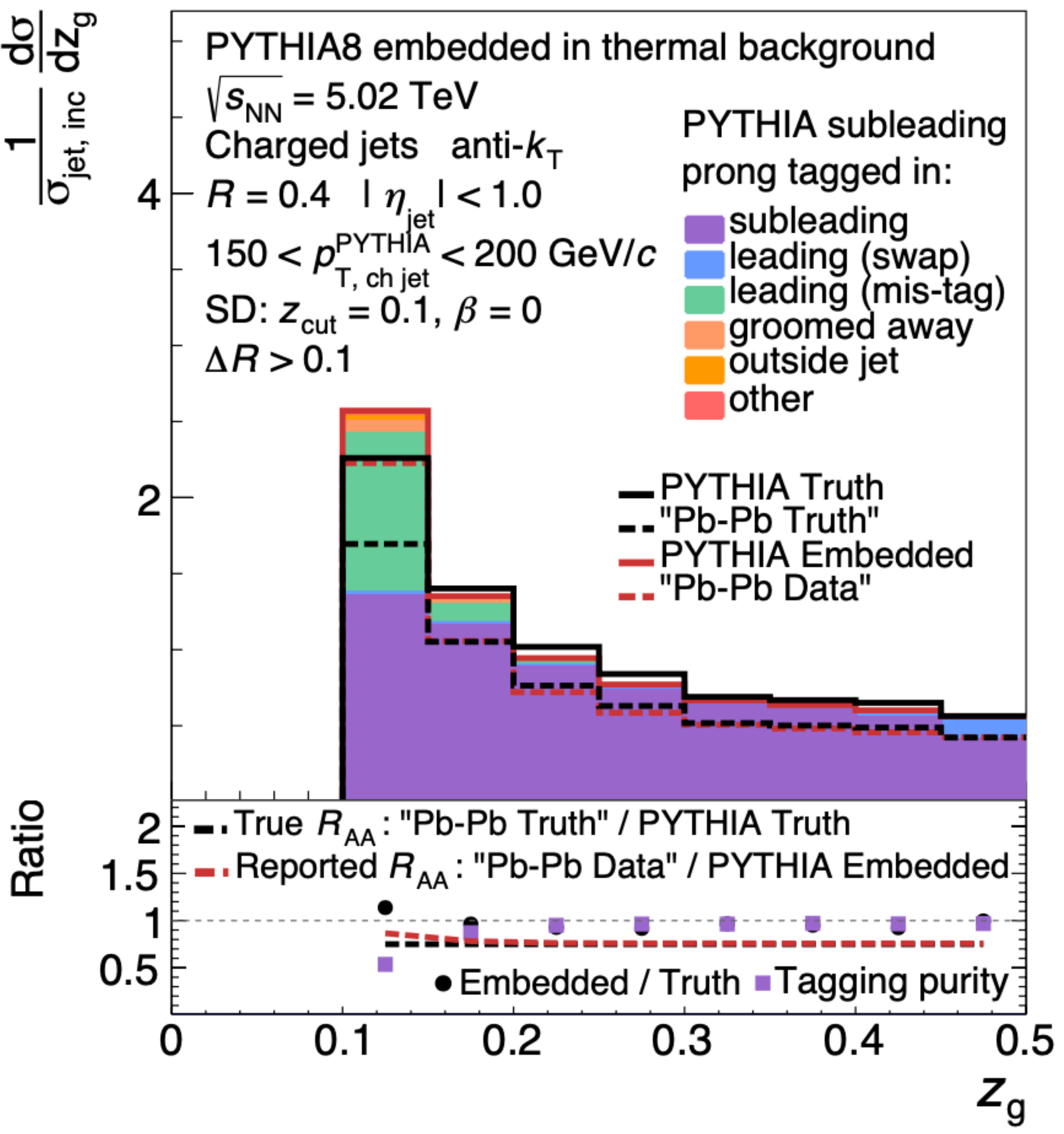}
\caption{Simple model showing that the presence of mistagged splittings
can induce an artificial shape in the \zg{} ratio, unrelated to jet quenching.
Here, the normalization due to the $R_{g}$ selection (denoted by $\Delta R$)
are taken in both numerator
and denominator to be from the PYTHIA distribution, in order to remove smearing
effects (but keep the suppression quantified by \RAA). Note that the momentum scale here is
taken from PYTHIA, whereas the experimental selection is a partially uncorrected
\PbPb{} scale.}
\label{fig:toy-zg}
\end{figure}

First, mistagged splittings dilute quenching effects, which can change
the shape of apparent modifications.
When comparing \PbPb{} data to an embedded reference, 
mistagged subleading prongs are not expected exhibit jet quenching,
since they arise from the combinatorial background.
Since the tagging purity varies with \zg{}, this means that
non-trivial changes to the shape of the \PbPb{}/\pp{} ratio 
can be induced. In particular, the tagging purity is low at small
values of \zg{}, and high at large values of \zg{}. 
To illustrate the impact of this, consider a simple
toy example for kinematics similar to the ALICE measurement with 
$\Delta R > 0.2$, shown in Fig. \ref{fig:toy-zg} left. 
Suppose that the true $\RAA$ induced by the \Rg{} cut is $0.5$, independent of \zg{}.
If we assume that mistagged splittings are unaffected by jet quenching,
then the observed AA distribution will be given by:

\[
P_{AA}(z_g)= \fmatched \RAA P_{pp}(z_g) + (1-\fmatched)P_{pp}(z_g),
\]
where \fmatched{} is the tagging purity. Note that as 
$\fmatched \rightarrow 1, P_{AA}(z_g) \rightarrow R_{AA} P_{pp}(z_g)$,
whereas if 
$\fmatched \rightarrow 0, P_{AA}(z_g) \rightarrow P_{pp}(z_g)$.
Since the tagging purity is low at small \zg{} and high at large \zg{}, 
this generically causes the observed \RAA{} to exhibit an apparent relative
suppression of symmetric splittings -- due entirely to background effects,
and unrelated to jet quenching. We note that the exact shape of 
the apparent relative suppression is model-dependent; there
are many model-dependent choices one could make which we do not pursue 
further here;\footnote{
(A) The shape of the true $R_{\mathrm{AA}}$ could be different -- it could for
    example even show enhanced suppression of asymmetric splittings.
(B) The mistagged splittings may exhibit a nontrivial correlation to \Rg, 
    and/or be affected by quenching.
    Consider the case of the true
    subleading prong being absorbed into the true leading prong, due to a
    large-angle local background fluctuation becoming the subleading prong.
    We then have 
    $z_{\mathrm{fake}}=p_{t,\mathrm{bkgd}}/\left(p_{t,\mathrm{bkgd}}+p_{t,\mathrm{lead},\mathrm{true}}+p_{t,\mathrm{sub},\mathrm{true}}\right)$.
    In AA, the true prongs undergo energy loss, which may shift the \zg{}
    distribution towards larger values relative to pp embedded 
    in a background.
(C) The purity depends on both the model of the background
    and the jet.
}
however the feature that the measured \RAA{} will exhibit a spurious
relative suppression emerges generically, independent of the details
of jet quenching, and depending only on the fact that the purity is 
low at small \zg{} and high and large \zg{}. 
Based on these considerations it is difficult to conclude
that symmetric splittings are more suppressed than asymmetric
splittings using the ALICE measurement alone.
The right panel of Fig. \ref{fig:toy-zg} shows a similar toy example corresponding
approximately to CMS kinematics, which suggests that dilution effects
are substantially smaller due to the higher purity at high \pt{}, 
but may still be significant.
Note that if one fully corrects the distributions via unfolding 
instead of performing detector-level embedding 
comparisons, one eliminates the susceptibility to dilution effects, since the response
matrix encodes appropriate corrections of any residual mistagged splittings to their true splittings.

Second, the magnitude of MC-based corrections (relevant to \cite{2020135227})
grows as the number of mistagged splittings grows. 
In Fig. \ref{fig:toy-zg} (left), the ratio ``Embedded/Truth'' gives an estimate of the
size of MC-based corrections one has to perform to compare \PbPb{} data 
to an embedded reference, and is on the order 100\%.
Note that the shape of this correction is correlated with
the experimentally observed modification. Moreover,
the distributions are effectively self-normalized, aside from the
suppression induced by the \Rg{} cut -- meaning that
small-\zg{} modification necessarily causes large-\zg{} modification.

\section{Conclusion}

We performed a set of basic studies on the behavior of various 
jet grooming algorithms in the presence of the large combinatorial 
background characteristic of heavy-ion collisions.
We found that such background and its region-to-region density fluctuations cause 
a significant number of splittings
to be incorrectly identified as a genuine structure of signal jets.
The robustness of groomers against this experimental challenge is an important 
criteria for their usage in
jet substructure measurements in heavy-ion collisions. 
We quantified the performance of grooming algorithms using the purity of the 
identified splittings differentially in both the jet momentum and 
individual substructure observables.
Our studies show that subleading prongs are prone to misidentification 
(lost, replaced by a background flux of particles, and 
thus often merged into the leading prong)
and that, in general, the contamination decreases (the groomer performance improves)
with increased $p_{T}$ of the jets.
We identified a set of grooming algorithms that perform relatively well; 
however, in our test setup, we found that groomers used in some of the existing heavy-ion
measurements result in a significant contamination of the reported distributions with 
false splittings. 
Since these background induced splits can generically mimic jet quenching effects,
future measurements at the LHC and RHIC 
will need to leverage the grooming algorithms that maximize the purity of the 
genuine splittings.
One of the important challenges will be to properly quantify the 
uncertainties in the reported quantities due to residual contamination effects.

The studies presented here ought to be extended to further explore the model-dependence 
of the background and the impact of jet fragmentation on the performance of
grooming algorithms.
Moreover, similar purity studies should be extended to any observable
where a substructure object must be tagged jet-by-jet, such as reclustered subjets.
Alternate experimental approaches, such as ensemble-based background subtraction
of mistagged splittings, should also be explored.
The groomers that we have considered can be refined and expanded --
a promising direction to explore may be to combine a high-purity 
groomer with an additional phase space selection (e.g. $\kappa,t_f$). 
Investigation of alternate reclustering algorithms
and iterative grooming techniques may also be promising, as ultimately one
wants to optimize a 
combination of the reclustering algorithm and grooming condition to 
construct splittings that are both robust to mistagging and 
sensitive to relevant physics (and calculable).
This, of course, calls for further theoretical guidance.

\section*{Acknowledgements}

We thank Peter Jacobs,  Marco van Leeuwen, Felix Ringer, and Nima Zardoshti 
for helpful discussions.
    This work was supported by the U.S. Department of Energy, Office of Science, 
    Office of Nuclear Physics, under the contract DE-AC02-05CH11231.

\addcontentsline{toc}{section}{References}
\bibliography{main}

\end{document}